\documentclass[letterpaper]{aa}
\usepackage{graphicx}
\usepackage{graphics}
\usepackage{amssymb}

\begin{document}
%%%%%%%%%%%%%%%%%%%

\title{A quasi-time-dependent radiative transfer model of OH\,104.9+2.4${}^\star$}

\author{D.\,Riechers\inst1{${}^,$}\inst2, 
Y.\,Balega\inst3,
T.\,Driebe\inst2,
K.-H.\,Hofmann\inst2, 
A. B. Men'shchikov\inst4,
V. I. Shenavrin\inst5,
and G.\,Weigelt\inst2}

   \institute{Max-Planck-Institut f\"ur Astronomie, K\"onigstuhl 17,
              D-69117 Heidelberg, Germany 
%              \email{riechers@mpia-hd.mpg.de}
              \and
              Max-Planck-Institut f\"ur Radioastronomie, Auf dem
              H\"ugel 69, D-53121 Bonn, Germany
              \and
              Special Astrophysical Observatory, Nizhnij Arkhyz,
              Zelenchuk district, 
%369167, 
Karachai-Cherkessian Republic, Russia
              \and
              Institute for Computational Astrophysics, Saint Mary's
              University, Halifax, NS B3H 3C3, Canada
              \and
              Crimean Astrophysical Observatory, Nauchny, 98409, Crimea, Ukraine
              }

\offprints{ D.~Riechers \\ \email{riechers@mpia-hd.mpg.de} \\
${}^\star$Based on data collected at the 6\,m BTA telescope 
of the Special Astrophysical Observatory}

  \date{Received 7 October 2004 / Accepted 8 February 2005}

\authorrunning{D.\,Riechers et al.}
\titlerunning{A quasi-time-dependent radiative transfer model of OH\,104.9+2.4}

\abstract{We investigate the pulsation-phase dependent properties of the circumstellar dust shell (CDS) of 
the OH/IR star \object{OH\,104.9+2.4} based on radiative transfer modeling
(RTM) using the code DUSTY. Our 
previous study concerning simultaneous modeling of the spectral energy distribution (SED) and near-infrared 
(NIR) visibilities (Riechers et al.\ \cite{rie04}) has now been extended by means of a more detailed analysis 
of the pulsation-phase dependence of the model parameters of \object{OH\,104.9+2.4}. In order to investigate 
the temporal variation in the spatial structure of the CDS, additional NIR
speckle interferometric observations in the $K'$ 
band were carried out with the 6\,m telescope of the Special Astrophysical
Observatory (SAO). At a wavelength of 
$\lambda = 2.12\,\mu$m the diffraction-limited resolution of 74\,mas was
attained. Several key parameters of our previous best-fitting model had to be adjusted in
order to be consistent with the newly extended amount of observational data. It was
found that a simple rescaling of the bolometric flux $F_{\rm bol}$ is 
not sufficient to take the variability of the source into account, as the change in optical depth $\tau$ over 
a full pulsation cycle is rather high. On the other hand, the impact of a change in effective
temperature $T_{\rm eff}$ on SED and visibility is rather small. However,
observations, as well as models for other AGB stars, show the necessity of
including a variation of $T_{\rm eff}$ with pulsation phase in the radiative
transfer models. Therefore, our new best-fitting model accounts for these changes. 

%-- techniques: image processing -- stars: late-type 
}

   \maketitle
%%%%%%%%%%%%%%%%%%%%%%%%%%%%%%%%%%%%%%%%%%%%%%%%%%%%%%%%%%%%%%%%%%%%%%%%%%%%%%%%%

%%%%%%%%%%%%%%%%%%%%%%%%%%%%%%%%%%%%%%%%%%%%%%%%%%%%%%%%%%%%%%%%%%%%%%%%%%%%%%%%%
   \section{Introduction}\label{intro}
%%%%%%%%%%%%%%%%%%%%%%%%%%%%%%%%%%%%%%%%%%%%%%%%%%%%%%%%%%%%%%%%%%%%%%%%%%%%%%%%%
\object{OH\,104.9+2.4} (\object{IRAS\,22177+5936}, \object{AFGL\,2885}, \object{NSV\, 25875}, 
\object{IRCO\,243}) is an OH/IR type II-A class star. Such objects are highly
evolved asymptotic giant branch 
(AGB) stars exhibiting dusty envelopes, which are believed to be the main source of dust grains 
in our Galaxy. While the peak of the SED is expected to be around 6 --
10\,$\mu$m for OH/IR 
stars, the SiO and Si$\rm O_2$ features at 9.7\,$\mu$m and 18\,$\mu$m are
usually found in absorption, 
indicating a high optical depth of the CDS. This is consistent with the high mass-loss rates 
($10^{-7} - 10^{-4}\,M_{\odot}/$yr, Habing \cite{hab96}) that have been measured for these long-period variables (LPV). A 
large number of observational constraints were found for this class of stars, leading to detailed 
models of the evolution of these objects and the properties of their CDS (see Riechers et al.\ 
\cite{rie04}, hereafter Paper I, and references therein). 
For \object{OH\,104.9+2.4}, an outflow velocity of $v_{\rm e} = 15$\,km/s was measured 
by te Lintel Hekkert et al.\ (\cite{TLH91}), and the absolute distance was determined to be 
$D = 2.38 \pm 0.24$\,kpc (Herman \& Habing \cite{her85}). From IR observations, we derived a 
pulsation period of $P = 1500 \pm 11\,$d, as well as a bolometric flux ranging
from $F_{\rm bol} = 0.7 \times 10^{-10}\,\rm{W/m^2}$ at minimum phase to 
$F_{\rm bol} = 2.3 \times 10^{-10}\,\rm{W/m^2}$ at maximum phase (see Paper I
for details). \\
\indent The purpose of the present paper is to discuss our improved best-fitting
radiative transfer model  of \object{OH\,104.9+2.4}. The paper is organized as follows: in Sect.\ 2, 
results of our new NIR speckle interferometric observations are presented. 
Sections 3.1 and 3.2 summarize the results of our best-fitting models from
Paper I (hereafter model M-I) and our new best-fitting model (hereafter model M-II). 
Section 3.3 tests the model M-II quantitatively by comparing it with the SED data collected in Paper I for 
different pulsation phases. Section 4 summarizes and evaluates the results and closes the 
discussion with an outlook on future research. 

%%%%%%%%%%%%%%%%%%%%%%%%%%%%%%%%%%%%%%%%%%%%%%%%%%%%%%%%%%%%%%%%%%%%%%%%%%%%%%%%%
   \section{Speckle interferometric observations and data reduction}\label{obsres}
%%%%%%%%%%%%%%%%%%%%%%%%%%%%%%%%%%%%%%%%%%%%%%%%%%%%%%%%%%%%%%%%%%%%%%%%%%%%%%%%%
In addition to our speckle interferometric observations in 2002, which were
presented in Paper I, $K'$-band speckle interferograms of \object{OH\,104.9+2.4} were obtained with the 
Russian 6\,m telescope of the Special Astrophysical Observatory on October 11, 2003. 
The data were recorded with our HAWAII speckle camera through an interference filter
with a center wavelength of 2.12\,$\mu$m and bandwidth of 0.21\,$\mu$m ($K^{\prime}$ band).
Additional speckle interferograms were taken for the unresolved reference star
2MASS J22191350+5949584. 
With a pixel size of 28.7\,mas and seeing of 2.6\,arcsec, 
4100\,object frames and 4600\,frames of the reference star were taken, each with an exposure time 
of 218\,ms. These interferograms were used to compensate for the speckle interferometry transfer function. 
The $K'$-band visibility function of \object{OH\,104.9+2.4} was derived from the speckle interferograms using 
the speckle interferometry method (Labeyrie \cite{l70}). The reconstructed two-dimensional visibility 
at 2.12\,$\mu$m and azimuthally averaged visibility profile are shown in Fig.\,1. 
Given the accuracy and resolution of our visibility measurements, 
we find that the diameter ratio between major and minor axes differs from
unity by less than 6\%, which justifies the assumption of spherical symmetry.\\
\indent From the pulsation period derived for \object{OH\,104.9+2.4} in Paper
I, the phase of observations presented in this paper is determined to be $\Phi^{2003} =
0.25$, as compared to the phase of the 1996 ISO spectrum ($\Phi^{\rm ISO} =
0.5$) and to the phase corresponding to the 2002 SAO visibility measurements
($\Phi^{2002} = 0$). Taking into account the uncertainties in the pulsation
period caused by the low number of photometric data points and the resulting
limitation to a rather simple fitting procedure, we estimate the total error
of these phases to be $\sim$0.02.

%%%%%%%%%%%%%%%%%%%%%%%%%%%%%%%%%
%%%% Fig.1: SAO 2D-vis + 1D-vis
%%%%%%%%%%%%%%%%%%%%%%%%%%%%%%%%%
\begin{figure*}
      \begin{center}
\begin{minipage}[t]{65mm}{
\resizebox{60mm}{!}{
\includegraphics[angle=0]{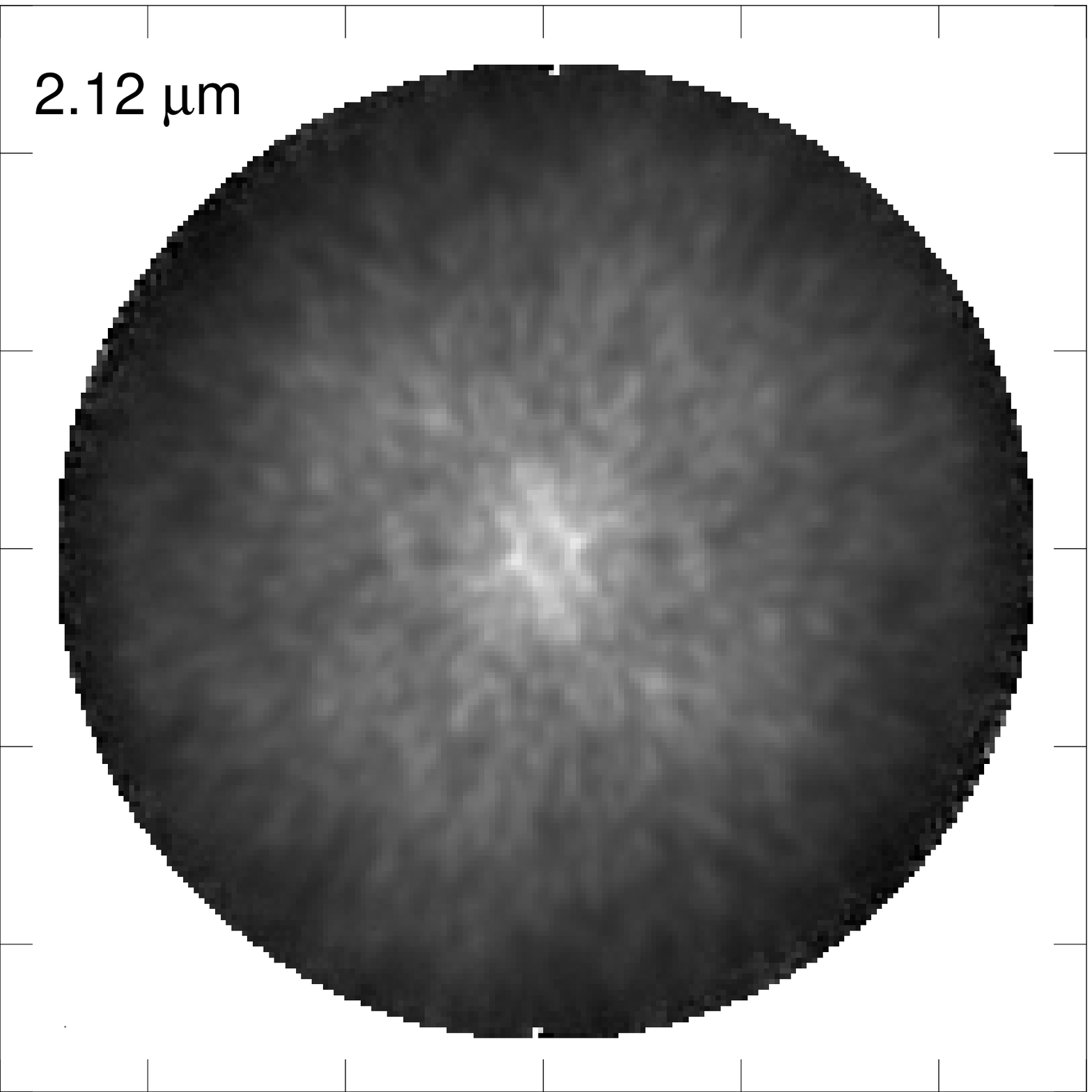}}
}
\end{minipage}
\begin{minipage}[t]{60mm}{
\resizebox{60mm}{!}{
\includegraphics[angle=0]{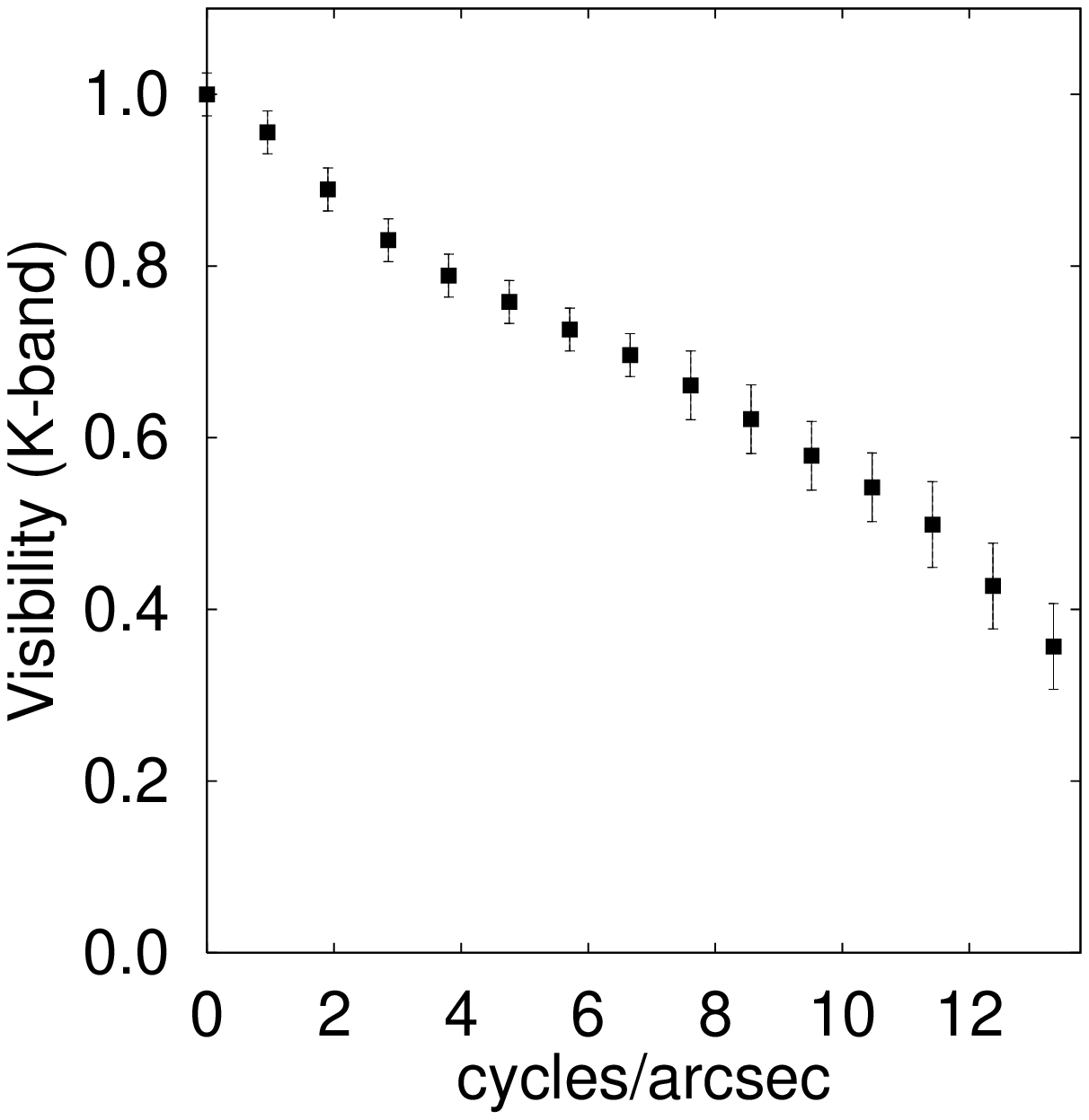}}
}
\end{minipage}
\begin{minipage}[t]{40mm}{
\vspace*{-50mm}

\caption{
Observations of \object{OH\,104.9+2.4} on October 11, 2003, with the SAO 6\,m telescope. 
\textbf{Left}: 2-dimensional visibility function of \object{OH\,104.9+2.4} at 2.12\,$\mu$m. 
\textbf{Right}: Azimuthally averaged visibility of \object{OH\,104.9+2.4} in
the $K^{\prime}$ band. 
}
}
\end{minipage}\hfill
      \end{center}
   \label{f1}
\end{figure*}
%%%%%%%%%%%%%%%

%%%%%%%%%%%%%%%%%%%%%%%%%%%%%%%%%%%%%%%%%%%%%%%%%%%%%%%%%%%%%%%%%%%%%%%%%%%%%%%%%
   \section{Dust shell models}\label{models}
%%%%%%%%%%%%%%%%%%%%%%%%%%%%%%%%%%%%%%%%%%%%%%%%%%%%%%%%%%%%%%%%%%%%%%%%%%%%%%%%%

%%%%%%%%%%%%%%%%%%%%%%%%%%%%%%%%%%%%%%%%%%%%%%%%%%%%%
\subsection{Previous results: Model M-I}
%%%%%%%%%%%%%%%%%%%%%%%%%%%%%%%%%%%%%%%%%%%%%%%%%%%%%

As reported in Paper I, for the previous analysis of the radiative transfer in
the CDS of \object{OH\,104.9+2.4}, 
$\sim 10^6$ models were calculated. To provide a reference for our SED model
computations we used the 1996 ISO
spectrum, while information about the spatial structure was given by our
2002 SAO $K'$-band visibility. To simultaneously fit the SED and 
visibility data (i.e.\ with otherwise identical model parameters), $F_{\rm bol}$ was adjusted 
for the epoch of the visibility measurements with respect to the SED model, since the ISO spectrum and the 2002 
visibility data were taken at different pulsation phases of \object{OH\,104.9+2.4}. 
Therefore, M-I consists of an SED model where the bolometric flux was
calculated from the 1996 ISO spectrum and from a visibility model using a
bolometric flux for the 2002 SAO phase, which was derived from photometric
measurements by different authors (see Paper I, Sect.\ 3). With these
observational constraints and phase dependencies, we have found a model that 
fits the observations. The main 
parameters of M-I are summarized in Table \ref{tab-1}.

%%%%%%%%%%%%%%%%%%%%%%%%%%%%%%%%%%%%%%%%%%%%%%%%%%%%%%%%%
%%%% Tab.1: Parameters of the previous best-fitting model
%%%%%%%%%%%%%%%%%%%%%%%%%%%%%%%%%%%%%%%%%%%%%%%%%%%%%%%%%
\begin{table*}

\caption{
The derived and adopted physical parameters of \object{OH\,104.9+2.4} as provided by Model M-I. 
}

\label{tab-1}
\begin{center} 
\begin{tabular}{ l l }\hline\hline
Parameter & Value 
\\ \hline
Effective temperature (black-body) & $T_{\rm eff} = 2500 \pm 500\,$K \\
Temperature at inner CDS boundary & $T_{\rm in} = 1000 \pm 200\,$K \\
Density profile within the CDS & $\rho (r) \propto r^{-n}$ with $n = 2.0 \pm 0.1$ \\
Relative CDS thickness & $\frac{r_{\rm out}}{r_{\rm in}} = 10^p$ with $p = 5^{+\infty}_{-2}$ \\
Dust-grain distribution function & MRN, $n(a) \propto a^{-3.5}$ (Mathis, Rumpl, \& Nordsieck \cite{mrn77}) \\
Minimum grain size & $a_{\rm min} = 0.005 \pm 0.003\,\mu$m \\
Maximum grain size & $a_{\rm max} = 0.2 \pm 0.02\,\mu$m \\
Dust types & 95\,\% warm silicates from Ossenkopf, Henning \& Mathis
(\cite{oss92}) \\
& 5\,\% astronomical silicates from Draine \& Lee (\cite{DL84}) \\
Optical depth & $\tau_{0.55\,\mu\rm{m}} = 158 \pm 7$ \\
& $\tau_{2.2\,\mu\rm{m}} = 6.5 \pm 0.3$ \\
& $\tau_{9.7\,\mu\rm{m}} = 14.0 \pm 0.6$ \\
Radius & $R_{\star} \simeq 600\,R_{\odot} = 2.79\,$AU \\
Radius of inner CDS boundary & $R_{\rm in} = 9.1\,R_{\star} = 25.4\,$AU \\
Mass-loss rate & $\dot{M} = 2.18 \times 10^{-5}\,M_{\odot}/$yr \\
Bolometric flux & $F_{\rm bol} = 0.7 \times 10^{-10}\,{\rm W/m^2}$ \\
\hline
\end{tabular}
\end{center}
%${}^{\star}$: Mathis, Rumpl \& Nordsieck (\cite{mrn77}) 
%${}^{\star\star}$: Ossenkopf, Henning \& Mathis (\cite{oss92}), 
%${}^{\star\star\star}$: Draine \& Lee (\cite{DL84})
\vspace*{-5mm}

\end{table*}
%%%%%%%%%%%%%%%%%%%%%%%%%%%%%%%%%%%%%%%%%%%%%%%

%%%%%%%%%%%%%%%%%%%%%%%%%%%%%%%%%%%%%%%%%%%%%%%%%%%%%
\subsection{The new best-fitting model: M-II}
%%%%%%%%%%%%%%%%%%%%%%%%%%%%%%%%%%%%%%%%%%%%%%%%%%%%%

\begin{figure*}[t]
      \begin{center}
\resizebox{8.95cm}{!}{\includegraphics[angle=-90]{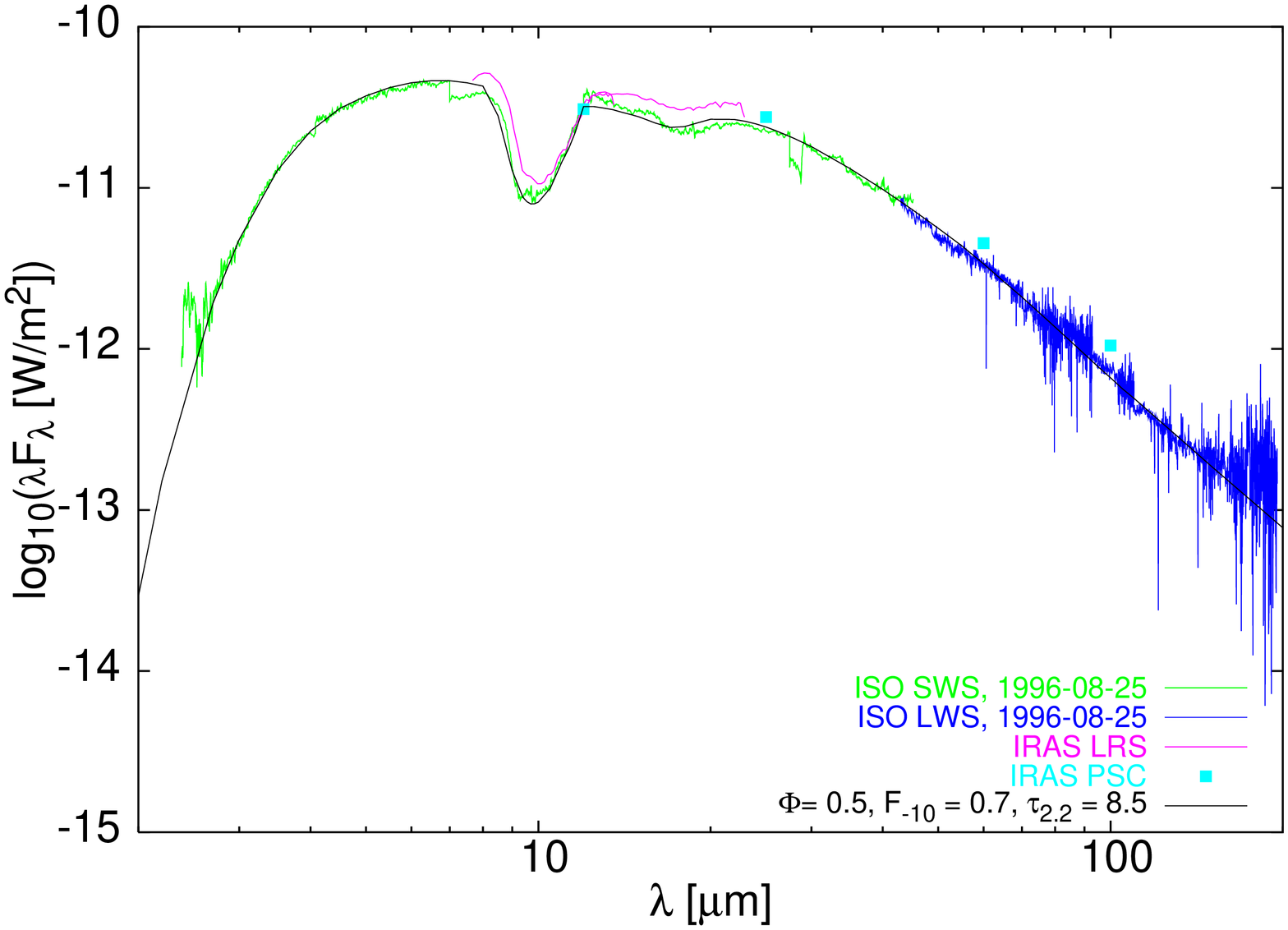}}
\resizebox{8.95cm}{!}{\includegraphics[angle=-90]{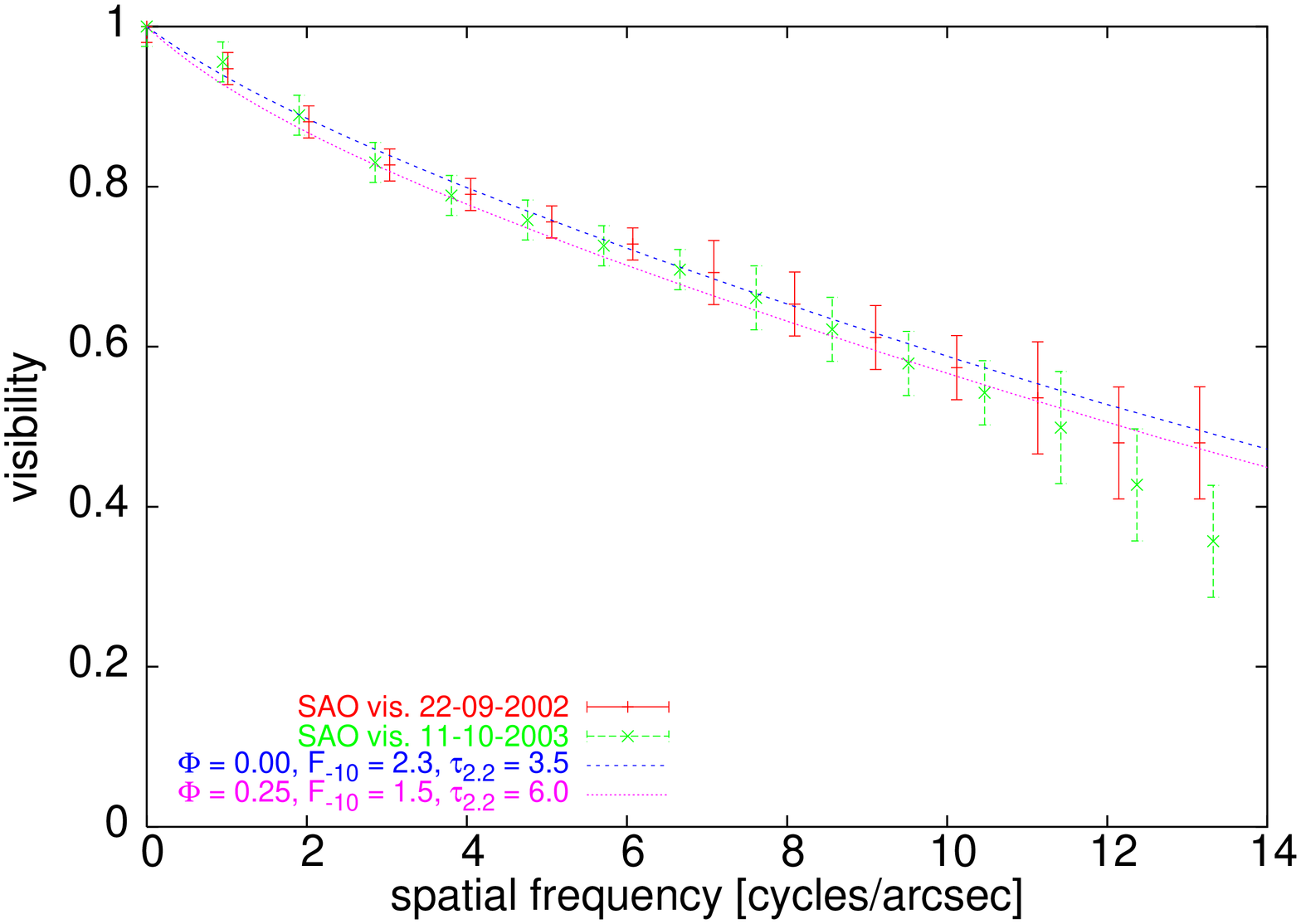}}
   \end{center}
\vspace*{-5mm}

\caption{
SED ({\bf left}) and 2.12\,$\mu$m visibility ({\bf right}) of \object{OH\,104.9+2.4} for our 
best-fitting model M-II. In order to take into account the different epochs of the ISO (1996) 
and SAO (2002 \& 2003) observations at 2.12\,$\mu$m, bolometric flux 
$F_{\rm bol}$, optical depth $\tau$, and  effective temperature $T_{\rm eff}$ had to be adjusted. 
The effective temperature of the central star is $T_{\rm eff}
= 2250\,$K for $\Phi = 0.5$, $T_{\rm eff} = 3150\,$K for $\Phi = 0$, and
$T_{\rm eff} = 2800\,$K for $\Phi = 0.25$. $F_{-10}$
denotes the bolometric flux in $10^{-10}\,{\rm W/m^2}$, while $\tau_{2.2}$ is
the optical depth at 2.2\,$\mu$m. 
}
   \label{f2}
\end{figure*}

Based on the new constraints given by the 2003 SAO visibility, some changes to M-I had to 
be introduced. The most significant change is that in addition to adjusting
$F_{\rm bol}$, a phase-dependent optical depth had to
be introduced in order to be able to 
explain the observations. This approach was recently suggested based on 
analysis of the depth of the 9.7\,$\mu$m feature for other OH/IR stars (see Suh \cite{S04}, and 
references therein). In addition, the new model considers variations in
the effective temperature of the central star, which is a less sensitive model
parameter than the optical depth. All results for M-II are 
summarized in Table \ref{tab-2}. The final model is shown in Fig.\,\ref{f2}. 

%%%%%%%%%%%%%%%%%%%%%%%%%%%%%%%%%%%%%%%%%%%%%%%%%%%%%
\subsubsection{Step 1: improved SED fit}
%%%%%%%%%%%%%%%%%%%%%%%%%%%%%%%%%%%%%%%%%%%%%%%%%%%%%

\begin{figure}
      \begin{center}
\resizebox{8.95cm}{!}{\includegraphics[angle=-90]{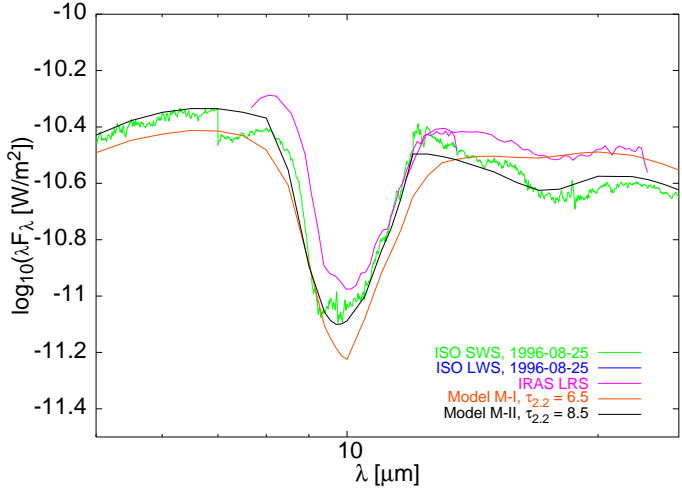}}
      \end{center}
\vspace*{-5mm}

\caption{SED of \object{OH\,104.9+2.4} in the mid-IR for models M-I (Paper I)
  and M-II (this work). Improvement in fitting the silicate absorption
  features is mainly due to the newly included dust composition. 
}
   \label{f3}
\vspace*{-3mm}

\end{figure}

Apart from the goal of obtaining a good fit for multiple pulsation phases, another aim of the ongoing research 
was to find a dust type that is capable of reproducing, in particular, the 18\,$\mu$m absorption feature 
better than the dust composition used in M-I. Therefore, we included in
our analysis several
dust types not inherent to the DUSTY code (Ivezi{\' c} et al.\ \cite{IE99}) in order to widen our parameter space 
grid. The only dust type that was found to reproduce the SED of \object{OH\,104.9+2.4} 
significantly better than the dust composition used in M-I is the cold silicates from Suh
(\cite{S99}). 
With this dust type, we have found a model with an Ivezi{\' c} et 
al.\ (\cite{IE99}) radiatively driven wind that gives a slightly better SED fit
than corresponding models with $\rho(r) \propto r^{-2}$. 
  In Fig.\,\ref{f3}, the SED in the 10\,$\mu$m regime for M-I
  and the final M-II are shown for comparison. \\
\indent The figure clearly illustrates how this new model is an improvement in
this region. At this point, we would like to note that
  the ISO SED of \object{OH\,104.9+2.4} shows some rather small emission features at
  33.3\,$\mu$m and 40.6\,$\mu$m, which are not considered in detail by our
  model even with the updated (as compared to M-I) dust composition. Suh (\cite{S02}) ascribes these
  features to crystalline silicates. However, it is difficult to extract
  information from those rather weak features and especially to positively identify
  them with specific dust components. Therefore, proper identification of
  these features may not be possible with current models. But, since these
  features are fairly weak, accounting for them in more detail would not
  significantly alter the conslusions of our study. \\
  \indent As we decided to use a model incorporating a
  radiatively driven wind, we had to comprehensively check our parameter sets
  to match the observed outflow velocities. The new best-fitting model
  results in an average terminal outflow velocity of $v_{\rm e}^{\rm M-II}
  = 16.8$\,km/s, which is in fair agreement with the observed outflow velocity
  of $v_{\rm e} = 15$\,km/s from te Lintel Hekkert et al.\ (\cite{TLH91}). \\
\indent In addition, the effective temperature was adjusted to $T_{\rm eff} =
2250\,$K for phase $\Phi = 0.5$. 
Considering the error bars found for M-I, this is only a minor adjustment. 

%%%%%%%%%%%%%%%%%%%%%%%%%%%%%%%%%%%%%%%%%%%
%%%% Fig.2: SED + vis., best-fitting model
%%%%%%%%%%%%%%%%%%%%%%%%%%%%%%%%%%%%%%%%%%%
%%%%%%%%%%%%%%%%%%%%%%%%%%%%%%%%%%%%%%%%%%%%%%%%

%%%%%%%%%%%%%%%%%%%%%%%%%%%%%%%%%%%%%%%%%%%%%%%%%%%%%%
%%%% Fig.3: SED - zoom on Si features
%%%%%%%%%%%%%%%%%%%%%%%%%%%%%%%%%%%%%%%%%%%%%%%%%%%%%%

%%%%%%%%%%%%%%%%%%%%%%%%%%%%%%%%%%%%%%%%%%%%%%%%%%%%%%

%%%%%%%%%%%%%%%%%%%%%%%%%%%%%%%%%%%%%%%%%%%%%%%%%%%%%%%%%
\subsubsection{Step 2: fitting multiple pulsation phases}
%%%%%%%%%%%%%%%%%%%%%%%%%%%%%%%%%%%%%%%%%%%%%%%%%%%%%%%%%

Comparing the 2002 and 2003 SAO visibility data, it is obvious that a simple rescaling of 
$F_{\rm bol}$ is insufficient for reproducing both observational
constraints (SED and visibility) with a single model. 
The visibility  scales with $\sqrt{F_{\rm bol}}$, and the difference in $F_{\rm bol}$ 
between the 2002 and the 2003 measurements is $\Delta F_{\rm bol} = 0.8 \times
10^{-10}\,{\rm W/m^2}$ (based on the assumption of a cosine-like pulsation
phase, see Paper I), 
corresponding to 35\% of $F_{\rm bol}^{\rm MAX}$. A difference of this order of magnitude 
cannot be derived from the SAO observations by merely rescaling the bolometric flux. Therefore, 
at least one additional parameter has to be changed to account for the observed variation of the 
visibility. To explain the observed change of depth for the 9.7\,$\mu$m
absorption feature of other OH/IR stars, different 
authors have suggested a change in optical depth with pulsation phase (see Suh \cite{S04}, and 
references therein). With both proper scaling of $F_{\rm bol}$ as a function of the pulsation phase 
($\Phi^{2002} = 0$, $F_{\rm bol}^{2002} = 2.3 \times 10^{-10}\,{\rm W/m^2}$; 
$\Phi^{2003} = 0.25$, $F_{\rm bol}^{2003} = 1.5 \times 10^{-10}\,{\rm W/m^2}$) and a linear 
adjustment of the optical depth ($\tau_{2.2\,\mu{\rm m}}^{\rm ISO} = 8.5$, 
$\tau_{2.2\,\mu{\rm m}}^{2002} = 3.5$, $\tau_{2.2\,\mu{\rm m}}^{2003} = 6.0$), the 1996 ISO 
SED, as well as the 2002 and 2003 SAO visibilities, can be reproduced by a
single model, as shown in Fig.\,\ref{f2}. \\
\indent Most detailed pulsation models for AGB stars predict an
increase of $T_{\rm eff}$ between minimum and maximum pulsation phases. For
Mira stars, Bessell et al.\ (\cite{BSW96}) predict the maximum in $T_{\rm
  eff}$ to be at $\Phi \simeq 0.8$. While radial pulsations of OH/IR stars
are driven by the same mechanism, it is difficult to assess whether the same
$\Phi$--$T_{\rm eff}$--relation is
true for these objects. However, our primary interest is to find a
suitable model that properly describes the relative changes between $\Phi = 0.5$, $\Phi
= 0$, and $\Phi = 0.25$. Based on Bessell et al.\ (\cite{BSW96}) and taking
into account that \object{OH\,104.9+2.4} is a cool star, we find $T_{\rm eff}
= 2250\,$K for $\Phi = 0.5$, $T_{\rm eff} = 3150\,$K for $\Phi = 0$, and
$T_{\rm eff} = 2800\,$K for $\Phi = 0.25$ are 
appropriate to account for the phase-dependent change of the central star's effective
temperature. 
Nevertheless, a change in
$T_{\rm eff}$ has only a small impact on SED and visibility and almost
the same effect as would a small adjustment of $\tau$ in the opposite direction. \\
\indent As can be seen from Table \ref{tab-2}, the maximum grain size $a_{\rm max}$ was 
increased for this model as compared to M-I. The new
value of $a_{\rm max} = 0.28\,\mu$m for M-II deviates from 
$a_{\rm max} = 0.20\,\mu$m for M-I by 4\,$\sigma$, but the change was necessary in order to enable 
us to scale the optical depth properly. This does not come as a surprise,
since, changing two different parameters in opposite directions may lead to nearly identical 
models, as pointed out in Paper I. Therefore, we were able to find different models of similarly good
quality in Paper I. The new 
best-fitting model M-II is much more constrained by observations than M-I, as the third 
observational constraint turned out to be essential in order to show the
necessity of adjusting the optical depth for different phases of \object{OH\,104.9+2.4}. \\
\indent Concerning the 2003 SAO visibility, M-II cannot reproduce the
observations beyond 12\,cycles/arcsec very well. This may be due to the large
observational errors close to the diffraction limit, which are possibly
slightly underestimated by the error bars shown here, as there is no
obvious physical reason why the visibility curvature should change 
as drastically as suggested by Fig.\,\ref{f2} in that regime between the 
dates of the two SAO observations. 

%%%%%%%%%%%%%%%%%%%%%%%%%%%%%%%%%%%%%%%%%%%%%%%%
%%%% Tab.2: Parameters of the best-fitting model
%%%%%%%%%%%%%%%%%%%%%%%%%%%%%%%%%%%%%%%%%%%%%%%%
\begin{table*}

\caption{
The derived and adopted physical parameters of \object{OH\,104.9+2.4} as provided by Model
M-II. Parameters that were treated as not time-dependent are indicated with
${}^{\star}$. A distance of $D = 2.38 \pm 0.24\,$kpc (Herman \& Habing
\cite{her85}) and an outflow velocity of $v_{\rm e} = 15\,{\rm km s^{-1}}$ 
(te Lintel Hekkert et al.\ \cite{TLH91}) were assumed. 
}

\label{tab-2}
\begin{center} 
\begin{tabular}{ l l l c c c }\hline\hline
Parameter & & & $\Phi = 0.5$ & $\Phi = 0$ & $\Phi = 0.25$ \\
& & & (1996) & (2002) & (2003)
\\ \hline
Black-body effective temperature & $T_{\rm eff}$ & [K] & 2250 & 3150 & 2800 \\
Temperature at inner CDS boundary${}^{\star}$ & $T_{\rm in}$ & [K] & 1000 & & \\
%Density profile within the CDS${}^{\star}$ & & & radiatively & driven & wind${}^{\star\star}$ \\
Relative CDS thickness${}^{\star}$ & $\frac{r_{\rm out}}{r_{\rm in}}$ & & $10^5$ & & \\
%Dust grain distribution function${}^{\star}$ & & & MRN${}^{\star\star\star}$ & & \\
Minimum grain size${}^{\star}$ & $a_{\rm min}$ & [$\mu$m] & 0.005 & & \\
Maximum grain size${}^{\star}$ & $a_{\rm max}$ & [$\mu$m] & 0.28 & & \\
Dust-to-gas ratio${}^{\star}$ & $r_{\rm dg}$ & & 0.005 & & \\
Dust grain bulk density${}^{\star}$ & $r_{\rm s}$ & [g\,cm$^{-3}$] & 3.0 & & \\
%Dust type${}^{\star}$ & & & Suh (\cite{S99}) & cold & silicates \\
Optical depth & $\tau_{0.55\,\mu\rm{m}}$ & & 146.7 & 60.4 & 103.6 \\
& $\tau_{2.2\,\mu\rm{m}}$ & & 8.5 & 3.5 & 6.0 \\
& $\tau_{9.7\,\mu\rm{m}}$ & & 13.9 & 5.7 & 9.8 \\
Radius & $R_{\star}$ & [$R_{\odot}$] & 729 & 675 & 691 \\
& & [AU] & 3.39 & 3.14 & 3.21 \\ 
& & [mas] & 1.43 & 1.32 & 1.35 \\
Radius of inner CDS boundary & $R_{\rm in}$ & [$R_{\star}$] & 8.3 & 17.5 & 13.7 \\
& & [AU] & 28.3 & 55.0 & 44.0 \\ 
& & [mas] & 11.9 & 23.1 & 18.6 \\
%Radius of $I = 10^{-10} I_{\rm max}$ & $R_{\rm out}^{-10} = 5 \times 10^3\,R_{\star} = 1.4 \times 10^4\,$AU & & \\
%Outer radius of the model & $R_{\rm out} = 9.1 \times 10^5\,R_{\star} = 2.5 \times 10^6\,$AU & & \\
Mass-loss rate & $\dot{M}$ & [$10^{-5}\,M_{\odot}/$yr] & 3.09 & 5.69 & 5.17 \\
Bolometric flux & $F_{\rm bol}$ & [$10^{-10}\,{\rm W/m^2}$] & 0.7 & 2.3 & 1.5 \\
Luminosity & $L$ & [$10^4\,L_{\odot}$] & 1.23 & 4.03 & 2.63 \\
%Terminal outflow velocity & $v_{\rm e}$ & [kms$^{-1}$] & 7.7 & 26.5 & 16.3 \\
\hline
\end{tabular}
\begin{tabular}{ l l l}
Density profile within the CDS${}^{\star}$ & radiatively driven wind &
(Ivezi{\' c} et al.\ \cite{IE99})\\
Dust-grain distribution function${}^{\star}$ & MRN, $n(a) \propto a^{-3.5}$ & (Mathis, Rumpl, \&
Nordsieck \cite{mrn77}) \\
Dust type${}^{\star}$ & S99 cold silicates & (Suh \cite{S99}) \\
\hline
\end{tabular}
\end{center}
%${}^{\star}$: parameter not varied with phase, 
%${}^{\star\star}$: Ivezi{\' c} et al.\ (\cite{IE99}), 
%${}^{\star\star\star}$: Mathis, Rumpl \& Nordsieck (\cite{mrn77}): $n(a) \propto a^{-3.5}$ 
%${}^{\star}$: parameter variation includes broken power laws, superwind models
%and models with radiatively driven winds\\
%${}^{\star\star}$: parameter variation includes test of alternative grain size 
%distributions (see text)\\
\vspace*{-5mm}

\end{table*}
%%%%%%%%%%%%%%%%%%%%%%%%%%%%%%%%%%%%%%%%%%%%%%%

%%%%%%%%%%%%%%%%%%%%%%%%%%%%%%%%%%%%%%%%%%%%%%%%%%%%%%%%%%
\subsubsection{Further model results}
%%%%%%%%%%%%%%%%%%%%%%%%%%%%%%%%%%%%%%%%%%%%%%%%%%%%%%%%%%

\begin{figure*}
      \begin{center}
\resizebox{78mm}{!}{\includegraphics[angle=-90]{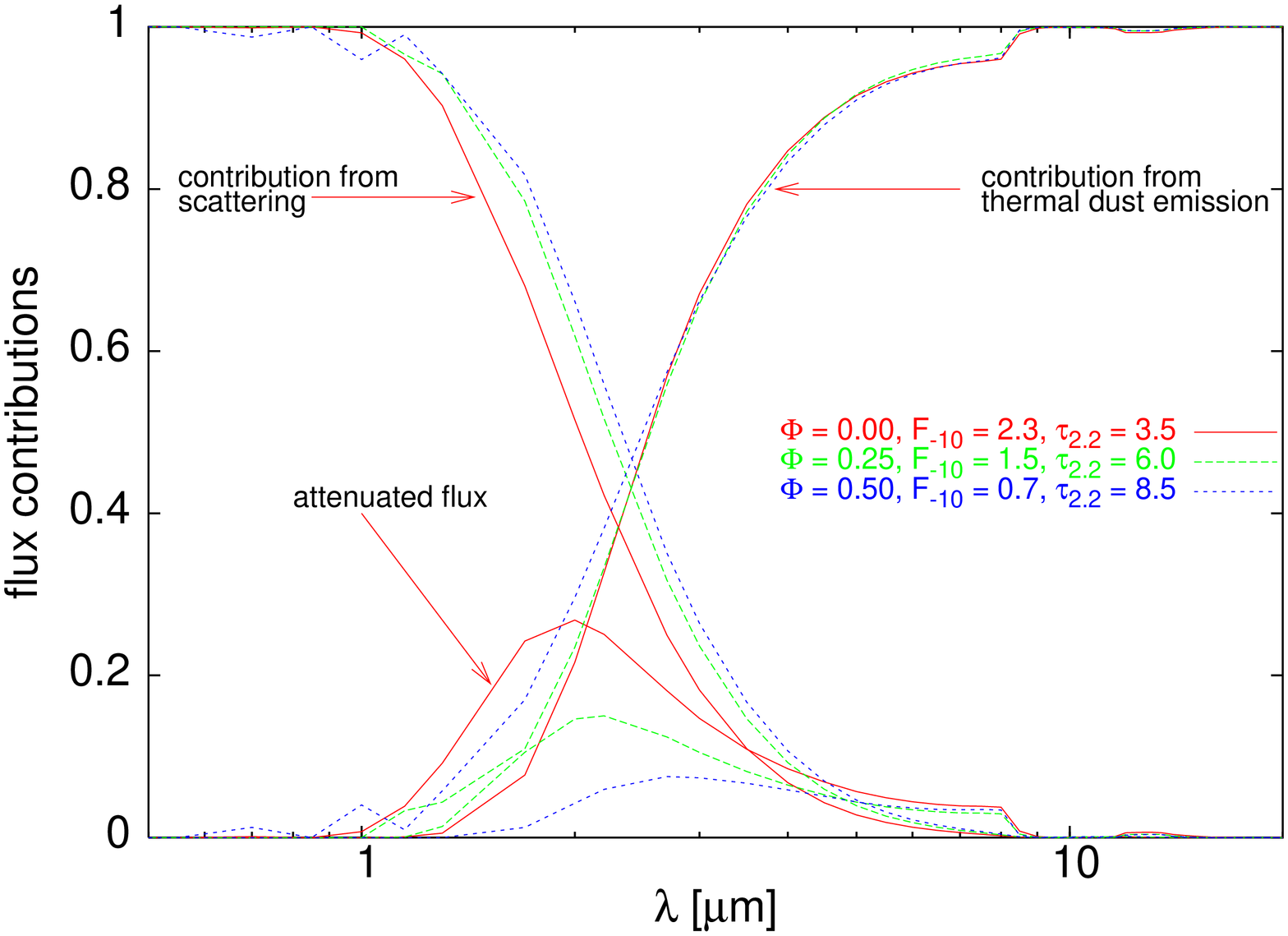}}
\resizebox{78mm}{!}{\includegraphics[angle=-90]{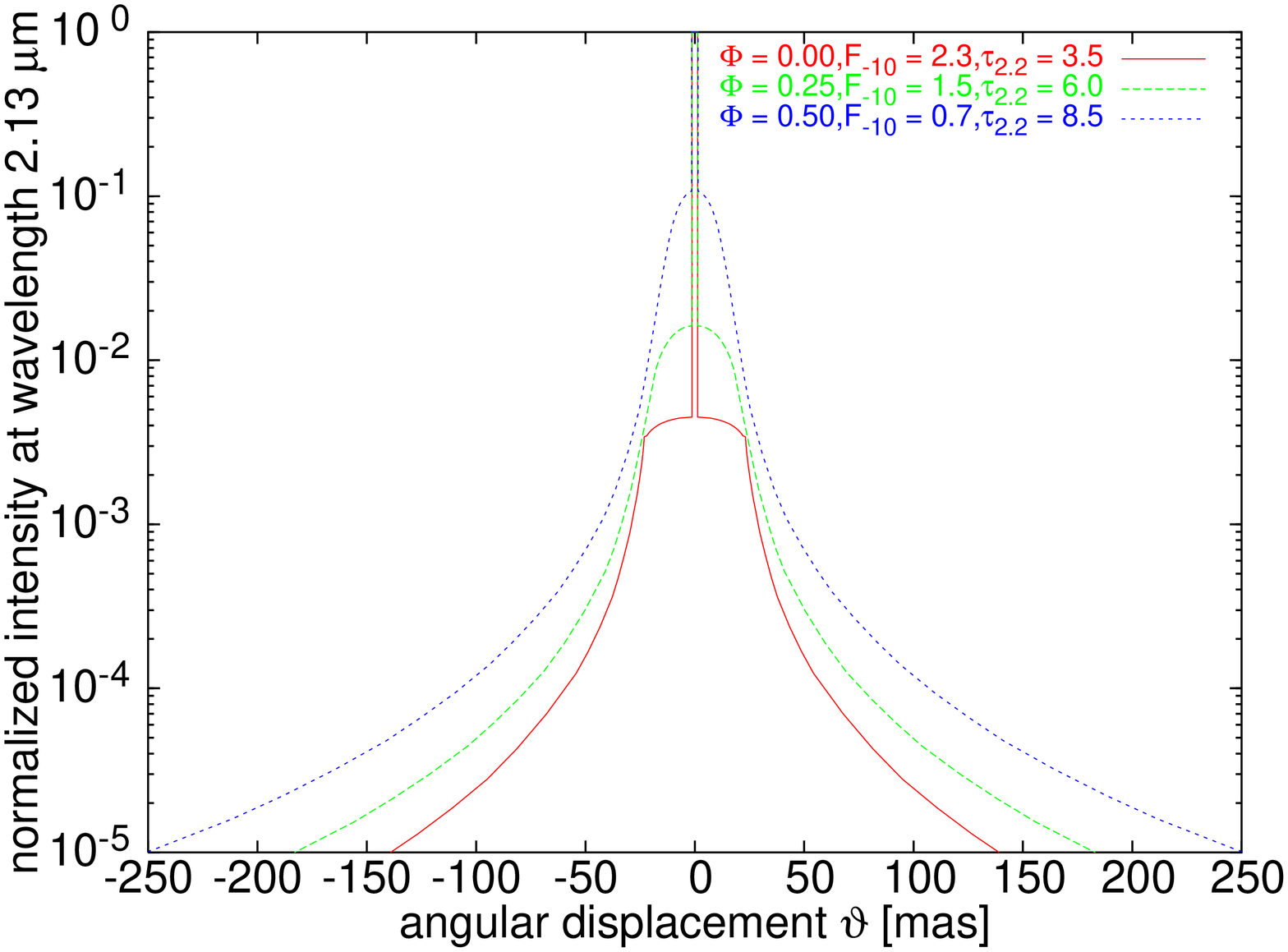}}
\resizebox{78mm}{!}{\includegraphics[angle=-90]{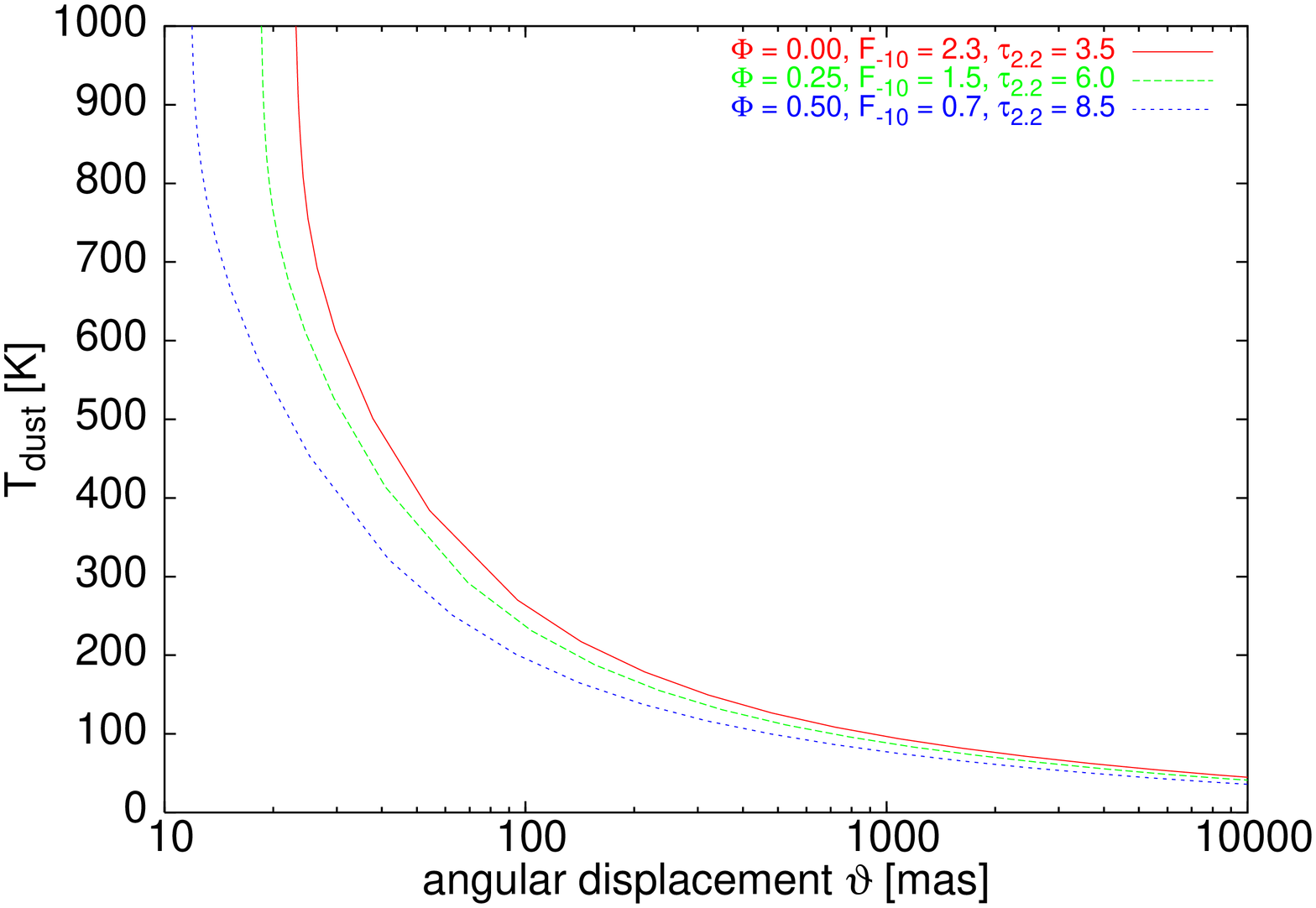}}
\resizebox{78mm}{!}{\includegraphics[angle=-90]{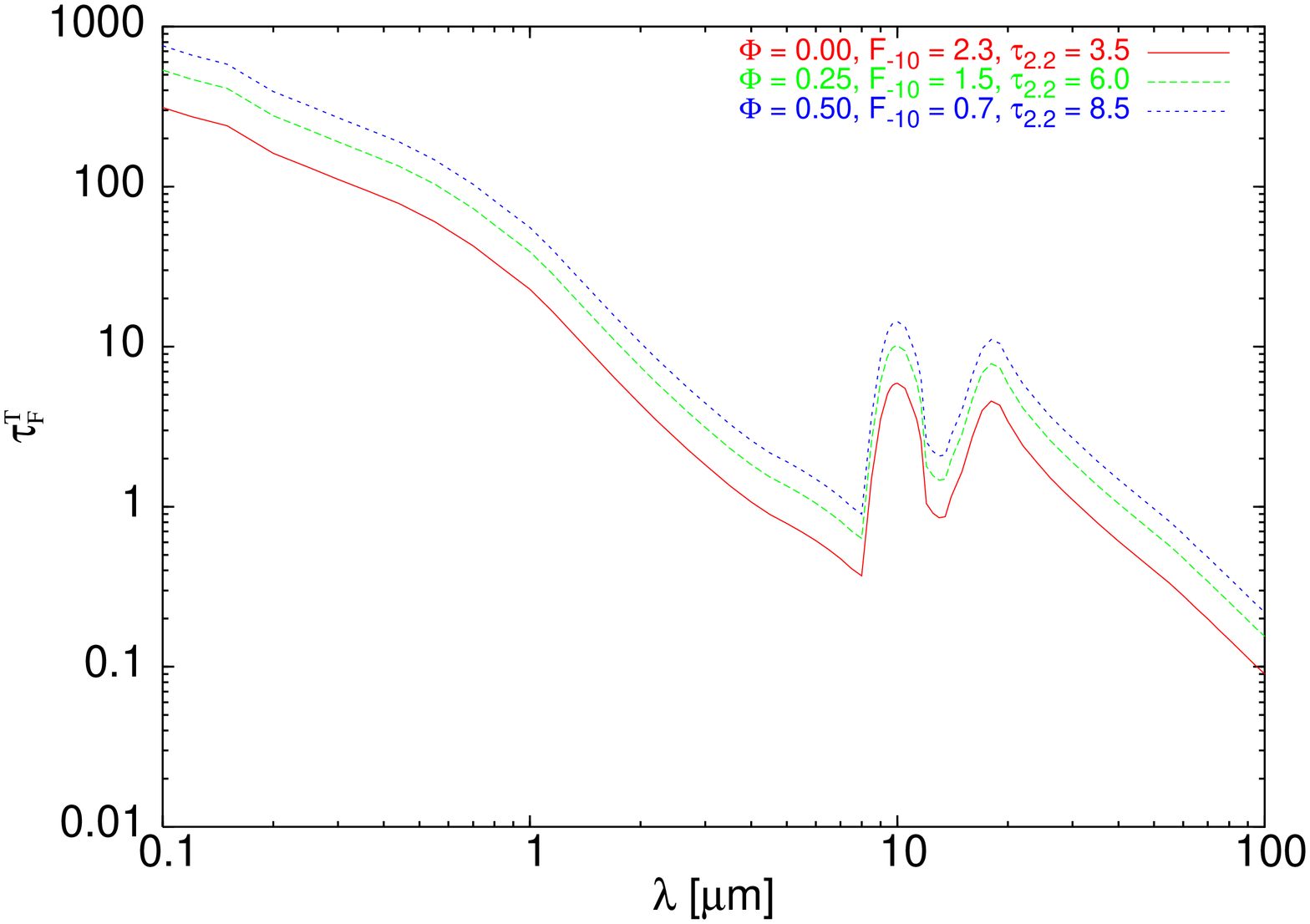}}
      \end{center}
\vspace*{-5mm}

\caption{
Modeling results of \object{OH\,104.9+2.4} for our best-fitting model M-II at the pulsation phases 
of the observations (see Table \ref{tab-2}). 
\textbf{Top left}: Fractional contributions to the total flux. The contribution from 
direct stellar light is small compared to the contribution of dust
scattering and thermal emission to the physical processes in the CDS (see
  Table \ref{tab-3}). 
\textbf{Top right}: The normalized intensity profile at 2.13\,$\mu$m. 
The sharp central peak corresponds to the central source of radiation. 
\textbf{Bottom left}: Dust temperature as a function of angular distance. 
The point where $T_{\rm dust} = 1000$\,K indicates the inner radius of the
dust shell. As the figure shows, due to the variability of the central source,
the inner dust shell boundary moves from 11.9\,mas to 23.1\,mas from minimum
to maximum pulsation phase. 
\textbf{Bottom right}: Wavelength depencence of the total optical depth.
}
   \label{f4}
\end{figure*}

The top left panel of Fig.\,\ref{f4} shows the fractional flux contributions of 
the emerging stellar radiation, scattered radiation, and thermal dust emission 
as a function of wavelength for model M-II. At 2.2\,$\mu$m, the flux 
is dominated by scattered light for all phases (see Table \ref{tab-3}). The contribution 
of direct stellar light (attenuated flux) rises from $\sim$6\% at minimum
phase to $\sim$25\% at maximum phase. \\
%
%%%%%%%%%%%%%%%%%%%%%%%%%%%%%%%%%%%%%%%%%%%%%%%
%%%% Fig.4: model profiles, best-fitting model
%%%%%%%%%%%%%%%%%%%%%%%%%%%%%%%%%%%%%%%%%%%%%%%
%%%%%%%%%%%%%%%%%%%%%%%%%%%%%%%%%%%%%%%%%%%%%%%%%%
%
\indent The top right panel of Fig.\,\ref{f4} shows the normalized intensity 
distribution at 2.13 $\mu$m as a function of angular distance. The barely resolved 
central peak corresponds to the central star. Here, the increase in the flux contribution 
of the central source towards maximum phase is reflected by the rise of
the central peak, as it does indeed become larger. 
In addition, the inner rim of the CDS itself moves outwards. \\
\indent The bottom left panel of Fig.\,\ref{f4} shows a plot of the dust temperature 
as a function of angular distance. The inner CDS radius $r_{\rm in}$ can be derived directly from the 
radial profile, as it is the point at which the dust reaches $T_{\rm in} = 1000$\,K. 
It corresponds to an angular radius of $\vartheta_{\rm in} = 11.9$\,mas for $\Phi = 0.5$ 
(for M-I: $\vartheta_{\rm in} = 10.5$\,mas), 
$\vartheta_{\rm in} = 18.6$\,mas for $\Phi = 0.25$, and $\vartheta_{\rm in} = 23.1$\,mas 
for $\Phi = 0$. This again shows that the inner rim of the CDS moves outwards
as the pulsation phase rises from minimum to maximum, mainly a
consequence of increasing $L_{\star}$. According to our model, the
change in the inner dust shell radius between minimum and maximum phases is
approximately a factor of two. \\
\indent The bottom right panel of Fig.\,\ref{f4} shows the total optical depth as a 
function of wavelength. As can be expected for the chosen chemical composition, local
maxima of $\tau_{\lambda }$ are found for the 
silicate absorption features at 9.7\,$\mu$m (SiO stretching vibrations) and 18\,$\mu$m 
(${\rm SiO_2}$ bending vibrations). The shape of the profile remains the same for all 
pulsation phases, but the optical depth is higher for all wavelengths towards
the minimum pulsation phase. \\
\indent The stellar radius drops by 8\% between minimum and maximum phases 
(from 729 to 675\,$R_{\odot}$), which is consistent with the Stefan-Boltzmann law. 
The radius at $\Phi = 0.5$ is approximately 18\% larger as compared to the value 
derived for M-I. As described in Paper I, the 
mass-loss rate $\dot{M}$ can be calculated using the observed outflow velocity $v_{\rm e} = 15$\,km/s 
and our derived luminosities for the different phases (see Table \ref{tab-2}). For minimum phase, 
$\dot{M} = 3.09 \times 10^{-5}\,M_{\odot}$/yr is obtained, which is roughly 50\% larger than the 
value found for M-I. For maximum phase, we derive $\dot{M} = 5.69 \times 10^{-5}\,M_{\odot}$/yr, 
which is by a factor of 1.8 larger than the mass-loss rate at minimum
pulsation phase. Heske et al.\ (\cite{hes90}) found $\dot{M} = 5.58 \times
10^{-5}\,M_{\odot}$/yr, which is in good agreement with our result.

%%%%%%%%%%%%%%%%%%%%%%%%%%%%%%%%%%%%%%%%%%
%%%% Tab.3: Flux contributions at 2.12 mu
%%%%%%%%%%%%%%%%%%%%%%%%%%%%%%%%%%%%%%%%%%
\begin{table}

\caption{
The contributions to the total flux at 2.2\,$\mu$m for
different pulsation phases $\Phi$ as given by Model M-II. 
}

\label{tab-3}
\begin{center} 
\begin{tabular}{ l l l l }\hline\hline
%& & phase & \\
component & $\Phi = 0.5$ & $\Phi = 0$ & $\Phi = 0.25$ 
\\ \hline
attenuated & 5.9\% & 25.1\%& 15.0\% \\
scattering & 55.9\% & 42.2\% & 51.7\% \\
thermal & 38.2\% & 32.7\% & 33.3\% \\
\hline
\end{tabular}
\end{center}
\vspace*{-5mm}

\end{table}

%%%%%%%%%%%%%%%%%%%%%%%%%%%%%%%%%%%%%%%%%%%%%%%%%%%%%%%%%%%

%%%%%%%%%%%%%%%%%%%%%%%%%%%%%%%%%%%%%%%%%%%%%%%%%%%%%
\subsection{Quality of the SED modeling at different phases}
%%%%%%%%%%%%%%%%%%%%%%%%%%%%%%%%%%%%%%%%%%%%%%%%%%%%%

\begin{figure}
      \begin{center}
\resizebox{8.95cm}{!}{\includegraphics[angle=-90]{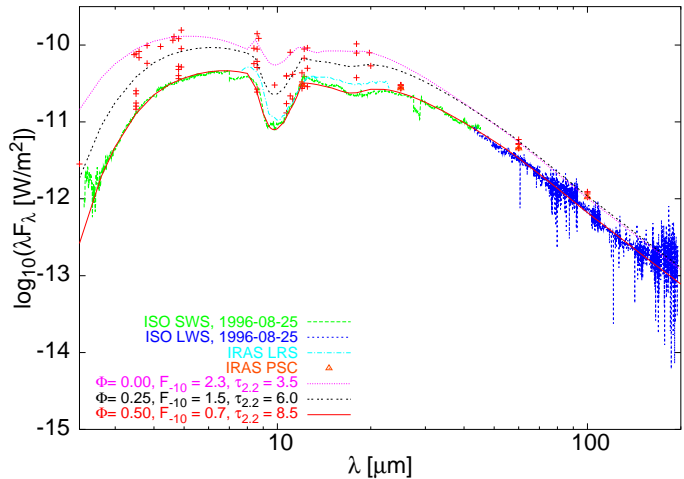}}
      \end{center}
\vspace*{-5mm}

\caption{
Variation of the SED with pulsation phase. All data collected in Paper I (see
Fig.\ 2 in Paper I and references therein) are plotted in 
order to test the quality of the SED fit of M-II for the 2002/2003 visibility phases. 
}
   \label{f5}
\vspace*{-3mm}

\end{figure}

In Sect.\ 3.2 it was shown that model M-II is able to properly reproduce the ISO SED and the 
SAO visibilities. But, as argued in Paper I, the SED varies in overall 
flux as well as in shape, depending on the pulsation phase. Therefore, this has 
to be tested if M-II is to reproduce the SED of \object{OH\,104.9+2.4} at 
different pulsation phases. Unfortunately, only few photometry is available
besides the ISO spectrum at minimum phase; while for some data, the pulsation phase 
is uncertain, as the observing date is not specified in the corresponding
publications. Therefore, we decided to perform only a 
qualitative test on the data used in Paper I to derive the SED at
maximum phase. 
Figure \ref{f5} shows the results. The models at minimum and maximum phase 
properly constrain the range of observational data. Thus, Model M-II seems to correctly 
reproduce the change in shape and the scaling with $F_{\rm bol}$ of the SED of 
\object{OH\,104.9+2.4}. 

%%%%%%%%%%%%%%%%%%%%%%%%%%%%%%%%%%%%%%%%%%%%%%%%%%%%%%
%%%% Fig.5: variation of the SED with pulsation phase
%%%%%%%%%%%%%%%%%%%%%%%%%%%%%%%%%%%%%%%%%%%%%%%%%%%%%%

%%%%%%%%%%%%%%%%%%%%%%%%%%%%%%%%%%%%%%%%%%%%%%%%%%%%%%

%%%%%%%%%%%%%%%%%%%%%%%%%%%%%%%%%%%%%%%%%%%%%%%%%%%%%%%%%%%%%%%%%%%%%%%%%%%%%%%%%
   \section{Summary and conclusions}\label{conclusions}
%%%%%%%%%%%%%%%%%%%%%%%%%%%%%%%%%%%%%%%%%%%%%%%%%%%%%%%%%%%%%%%%%%%%%%%%%%%%%%%%%

We present an improved, quasi-time-dependent radiative transfer model for the
dust shell of the type II-A 
OH/IR star \object{OH\,104.9+2.4}. This model is constrained by recent 
speckle-interferometric observations at 2.12\,$\mu$m and by 
spectro-photometric observations. The speckle observations were carried out with the 
SAO 6\,m telescope at a diffraction-limited resolution of 74\,mas, which is 
sufficient for resolving the CDS at this wavelength. The reconstructed 
2-dimensional visibility shows no major deviation from spherical symmetry. As
a reference SED, we used the ISO SWS and LWS spectra ($\lambda = 2.38 -
196\,\mu$m) from 1996, which correspond
to minimum pulsation phase, as well as additional photometric data at
different phases (see Paper I for details). \\
\indent Our simultaneous modeling of visibilities at different pulsation 
phases and of the ISO SED of \object{OH\,104.9+2.4} lead to modifications of our 
previous best-fitting model presented in Paper I. The new results demonstrate the need to 
consider changes in bolometric flux, as well as in optical depth and central
star effective temperature with pulsation phase, in order to obtain more reliable results. \\
\indent The final best-fitting model of this study leads to prediction of a 
temperature at the inner rim of the dust shell of $T_{\rm in} = 1000\,$K. The dust 
shell has a relative thickness of $10^5$, while the density decreases outwards 
according to a radiatively driven wind as
described in Ivezi{\' c} et al.\ (\cite{IE99}). The dust 
grains follow a standard MRN distribution with grain sizes $a = 0.005 - 0.28\,\mu$m, 
while the optical constants of the dust itself are represented best by the
cold silicates from Suh (\cite{S99}). The optical depth varies between minimum and maximum 
pulsation phase from $\tau_{2.2\,\mu{\rm m}} = 8.5$ to $\tau_{2.2\,\mu{\rm m}} = 3.5$, 
while the bolometric flux changes from $F_{\rm bol} = 0.7 \times 10^{-10}\,{\rm W/m^2}$ 
to $F_{\rm bol} = 2.3 \times 10^{-10}\,{\rm W/m^2}$. \\
\indent Based on pulsation models
for Mira stars (Bessell et al.\ \cite{BSW96}), variation of the effective
temperature of the central star with pulsation phase is taken into
account. For the cool star \object{OH\,104.9+2.4}, the best-fitting model
predicts a change from $T_{\rm eff} = 2250\,$K to $T_{\rm eff} = 3150\,$K
between minimum and maximum pulsation phases. While a change in $T_{\rm eff}$
shows only a minor impact on SED and/or visibility, it leads to a prediction of
the change in the stellar radius (from $R_{\star} = 729\,R_{\odot}$ at minimum
phase to $R_{\star} = 675\,R_{\odot}$ at maximum phase) for our best-fitting model, which is
in agreement with observations of similar objects (see, e.g., Bessell et al.\
\cite{BSW96}). 
The inner rim of the dust shell moves outwards 
by roughly a factor of two from minimum to maximum phase ($R_{\rm in} = 28.3 - 55.0\,$AU), while an 
increase of the mass-loss rate from $\dot{M} = 3.09 \times 10^{-5}\,M_{\odot}/$yr to 
$\dot{M} = 5.69 \times 10^{-5}\,M_{\odot}/$yr is found. Several observational
constraints (1996 ISO SED, 2002/2003 SAO visibilities, IR photometry at
different epochs) are represented very well by this model. Only 
the 2003 visibility at high spatial frequencies shows deviation from the 
model, which may be due to underestimation of observational errors. \\
\indent In the future, it would be very important to obtain $K'$-band visibility 
measurements at pulsation phase $\Phi = 0.5$ in order to check the validity of the 
phase-dependent model, although conclusions may be limited due to possible
cycle-to-cycle variations. Additional measurements for intermediate phases would
help to check if the optical depth really scales linearly with the pulsation amplitude or 
if a more complex approach has to be considered. 
Likewise, 
it is still necessary to carry out more photometric measurements in order to pin 
down the pulsation period with higher precision. Furthermore, visibilities at 
other wavelengths (probably $J$ or $H$, which poses a challenge due to the detection 
limit) would help to improve the modeling and would give 
more detailed information on the spatial structure of the dust shell. 

%%%%%%%%%%%%%%%%%%%%%%%%%%%%%%%%%%%%%%%%%%%%%%%%%%%%%%%%%%%%%%%%%%%%%%%%%%%%%%%%%
%%%%%%%%%%%%%%%%%%%%%%%%%%%%%%%%%%%%%%%%%%%%%%%%%%%%%%%%%%%%%%%%%%%%%%%%%%%%%%%%%
%%%%%%%%%%%%%%%%%%%%%%%%%
\begin{acknowledgements}
%%%%%%%%%%%%%%%%%%%%%%%%%
We thank the Infrared Space Observatory (ISO) operators at the European Space Agency 
(ESA) for providing the SED data. This research has made use of the SIMBAD database operated 
by CDS in Strasbourg as well as the Gezari catalogue published by the National Aeronautics and Space 
Administration (NASA), and the NASA Astrophysics Data System (ADS) operated by NASA. This publication 
makes use of data products from the Two Micron All Sky Survey (2MASS), which is a joint project of the 
University of Massachusetts and the Infrared Processing and Analysis Center/California Institute of 
Technology, funded by NASA and the National Science Foundation. A.\,B.\,M.\
acknowledges support from the Natural Sciences and Engineering Research
Council of Canada (NSERC). Finally, we thank the anonymous referee for
helpful comments that stimulated further improvements in the manuscript.
%%%%%%%%%%%%%%%%%%%%%%%%%
\end{acknowledgements}
%%%%%%%%%%%%%%%%%%%%%%%%%

%%%%%%%%%%%%%%%%%%%%%%%%%

%%%%%%%%%%%%%%%%%%%%%%%%%
\end{document}